\documentstyle[twoside,fleqn,espcrc2,epsfig]{article}

\newcommand{\AmS}{{\protect\the\textfont2
  A\kern-.1667em\lower.5ex\hbox{M}\kern-.125emS}}

\def    \np     #1#2#3{{\it Nucl. Phys.} {\bf #1}(19#2)#3}
\def    \pl     #1#2#3{{\it Phys. Lett.} {\bf #1}(19#2)#3}
\def    \prl    #1#2#3{{\it Phys. Rev. Lett.} {\bf #1}(19#2)#3}
\def    \pr     #1#2#3{{\it Phys. Rev.} {\bf #1}(19#2)#3}

\def    \eup    #1#2#3{{\it Eur. Phys. J.} {\bf #1}(#2)#3}
\def    \hepph  #1 {{\tt hep-ph/#1}}
\def    \hepex  #1 {{\tt hep-ex/#1}}

\def\bentarrow{\:\raisebox{1.3ex}{\rlap{$\vert$}}\!\longrightarrow}
\def\beq{\begin{equation}}
\def\eeq{\end{equation}}
\def\beqn{\begin{eqnarray}}
\def\eeqn{\end{eqnarray}}
\def\sss{\scriptscriptstyle}

\def\as{\alpha_{\sss S}}
\def\epem{e^+e^-}

\begin{document}

\twocolumn[
\vskip -3cm
~

\flushright{
        \begin{minipage}{3.2cm}
        CERN-TH/2000-289\hfill \\
        hep-ph/0010044\hfill \\
        \end{minipage}        }

\vskip 5.0em
\flushleft{\Large Higher-order QCD corrections to 
$\gamma^*\gamma^*\to hadrons$}~\footnotemark
\vskip\baselineskip
\flushleft{Stefano Frixione$^{\rm a}$}
\vskip\baselineskip
\flushleft{$^{\rm a}$CERN, TH division, Geneva, Switzerland}

\begin{center}
\begin{minipage}{160mm}
                \parindent=10pt
{\small 
I illustrate the techniques and the results of the computation of 
order-$\as$ corrections to the production of hadrons in the 
collision of two virtual photons, that originate from the 
incoming leptons at $\epem$ colliders.
}
                \par
                \end{minipage}
                \vskip 2pc \par
\end{center}
]
\footnotetext{Talk given at QCD00, 6-12 July 2000, Montpellier, F.}


\section{Introduction}

In the study of scattering phenomena in the limit of very large
energies at fixed momentum transfer, one must face the fact that the
two most promising probes of BFKL physics, namely small-$x$ effects 
in DIS and large-rapidity-gap events in dijet production at hadron
colliders, are inherently dependent upon long-distance physics, due 
to the presence of hadrons in the initial state; this poses serious
difficulties on the possibility of cleanly extracting signals of BFKL 
physics from data.

To overcome these difficulties a gedanken experiment has been 
proposed~\cite{Onium}, where two quarkonium states collide; the
transverse size of the quarkonium is small enough to allow the
perturbative computation of its wave function. Of course, since
quarkonium colliders are out of sight, one must devise some other
solutions; an increasing-popular one is the study of the process
\beq
\gamma^* +\gamma^*\;\longrightarrow\; hadrons,
\label{processgg}
\eeq
at fixed photon virtualities $-q_i^2=Q_i^2>0$, and for large 
center-of-mass energies squared $W=(q_1+q_2)^2$; here, we denote 
with $q_i$ the momenta of the photons. The virtual photons play 
the same role as the quarkonia; they are colourless, and their
virtualities control their transverse sizes, which are roughly proportional 
to $1/\sqrt{Q_i^2}$, thus allowing for a completely perturbative
treatment. Notice that the virtuality of the photon is therefore
physically equivalent to the (squared) mass of the quarkonium; however,
while the mass of the quarkonium is fixed by nature, the virtuality
of the photon can be controlled by the experimental setup.

The easiest way to access the process of eq.~(\ref{processgg}) is through 
the following reaction:
\beq
\begin{array}{rcl}
e^+ + e^- & \longrightarrow & e^+ + e^- + \underbrace{\gamma^* + \gamma^*} \\
 &  & \phantom{e^+ + e^- + \gamma^*\:}\bentarrow hadrons ;
\end{array}
\label{processee}
\eeq
namely, one considers $\epem$ collisions, selecting those events in
which the incoming leptons produce the photons through the elementary
$e^+e^-\gamma$ vertex, and the two photons eventually initiate the hard 
scattering that produces the hadrons. Data relevant to the process in 
eq.~(\ref{processee}) have been published by several experiments in
the past, and many more are expected in the near future by LEP
collaborations; it goes without saying that a linear collider would 
be the ideal place where to look at such processes.

It is clear that the process in eq.~(\ref{processee}) is non physical; 
rather, it has to be understood as a shorthand notation for a subset 
of Feynman diagrams contributing to the process that is actually observed:
\beq
e^+ +e^-\,\longrightarrow\,
e^+ +e^- + hadrons.
\label{fullproc}
\eeq
Other contributions to the process in eq.~(\ref{fullproc}) are, 
for example, those
in which the incoming $\epem$ pair annihilates into a photon or a $Z$,
eventually producing hadrons and a lepton pair, or those in which 
one (or both) of the two photons in eq.~(\ref{processee}) is replaced
by a $Z$. However, it is not difficult to devise a set of cuts such
that the process in eq.~(\ref{processee}) gives the only non-negligible
contribution to the process in eq.~(\ref{fullproc}). One can tag 
{\em both} the outgoing leptons, and retain only those events in
which the scattering angles of the leptons are small: in such a way,
the contamination due to annihilation processes is safely negligible.
Furthermore, small-angle tagging also guarantees that the photon
virtualities are never too large (at LEP2, one typically measures
$Q_i^2={\cal O}(10$~GeV$^2$); therefore, the contribution from processes
in which a photon is replaced by a $Z$ is also negligible.
Thus, it is not difficult to extract the cross section of the 
process \mbox{$\gamma^*\gamma^* \to hadrons$} from the data relevant 
to the process in eq.~(\ref{fullproc}).
\begin{figure}[thb]
\vspace{9pt}
\epsfig{figure=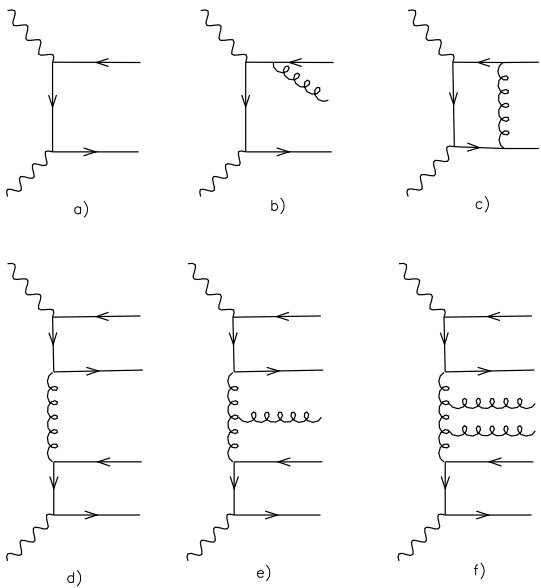,width=0.46\textwidth,clip=}
\caption{Sample of diagrams contributing to the production of hadrons
in two-photon collisions.}
\label{fig:diagr}
\end{figure}

Eventually, the data are compared to the theoretical predictions
obtained by using BFKL equation. Some of the diagrams contributing to
the process in eq.~(\ref{processgg}) are depicted in
fig.~\ref{fig:diagr}; diagrams d), e) and f) (plus all the diagrams
obtained by emitting more and more gluons from the $t$-channel gluon)
are included in BFKL dynamics. In the large-$W$ limit, diagrams with a
$t$-channel quark exchange, such as a), b) and c), are expected to give
a cross section behaving as
\beq
\sigma_{\gamma^*\gamma^*}\,\sim\,1/W,
\label{xsec}
\eeq
while diagrams relevant to BFKL physics, such as d), e) and f), are
expected to give
\beq
\sigma_{\gamma^*\gamma^*}^{\sss BFKL}\,\sim\,
1+\sum_{j=1}^\infty a_j (\as L)^j + {\cal O}(\as(\as L)^j),
\label{BFKLxsec}
\eeq
where \mbox{$L=\log(W/\mu_{\sss\rm W}^2)$} is a ``large'' logarithm, and 
all subleading logarithmic terms are indicated with 
${\cal O}(\as(\as L)^j)$; the quantity $\mu_{\sss\rm W}^2$ is a
mass scale squared, of the order of the photon virtualities.
By comparing eqs.~(\ref{xsec}) and (\ref{BFKLxsec}), it is clear that
the latter will dominate over the former in the asymptotic energy region
$W\to\infty$. However, at current collider energies,
$\sigma_{\gamma^*\gamma^*}$ is not safely negligible, and must be taken
into proper account when comparing theory and data. For this reason,
one usually {\it subtracts} the theoretical predictions for 
$\sigma_{\gamma^*\gamma^*}$ from data, and then compares the results
obtained in this way to the predictions for 
$\sigma_{\gamma^*\gamma^*}^{\sss BFKL}$. Unfortunately,
at present only the leading order contribution (diagram a) in 
fig.~\ref{fig:diagr}) to $\sigma_{\gamma^*\gamma^*}$ has been considered. 
This amounts to say that diagrams such as b) and c) have been neglected 
so far; these diagrams represent the first non-trivial QCD corrections 
to the process in eq.~(\ref{processgg}). We will denote these
contributions as next-to-leading order (NLO) corrections, although
effectively of leading order in $\as$. The aim of this work, which
presents preliminary results from ref.~\cite{CDFT}, is to report the
computation of these NLO corrections.

\section{Computation of NLO corrections}

The computation of the NLO corrections to a hard scattering process is
by now a rather standard procedure, since algorithms exist that are
universal (that is, process independent), and applicable to any number
of final state partons. The role of these algorithms is to combine
in a physically sensible way the virtual and the real contributions, 
that are unphysical and divergent upon loop and phase-space integrations.
The information on the hard process basically enter 
only in the computation of the matrix elements. In our case,
one needs to compute the amplitude of the process
\mbox{$\epem\to\epem q\bar{q}$} at one loop, and of the process
\mbox{$\epem\to\epem q\bar{q}g$} at the tree level. Fortunately, these
results are easily obtained from existing literature (notice that
we work with massless flavours; being mainly interested in the region of 
$W$ not close to the heavy-quark threshold, our approximation is sufficiently
good). As far as the one-loop amplitude is concerned, we can derive
it from the one-loop amplitude relevant to the process
\mbox{$\epem\to q_a\bar{q}_a q_b\bar{q}_b$} given in ref.~\cite{BDK}.
For the tree-level amplitude, we can use the results of ref.~\cite{DKS}
relevant to the process \mbox{$q\bar{q}\to Z^*Z^*g
\to l_a\bar{l}_a l_b\bar{l}_b g$}, with a suitable crossing and 
inserting the electromagnetic coupling factors instead of the 
electroweak ones; the fact that the $Z$ has an axial coupling is easily
dealt with, since the results of ref.~\cite{DKS} are given in terms
of helicity amplitudes.

\begin{figure}[thb]
\vspace{9pt}
\epsfig{figure=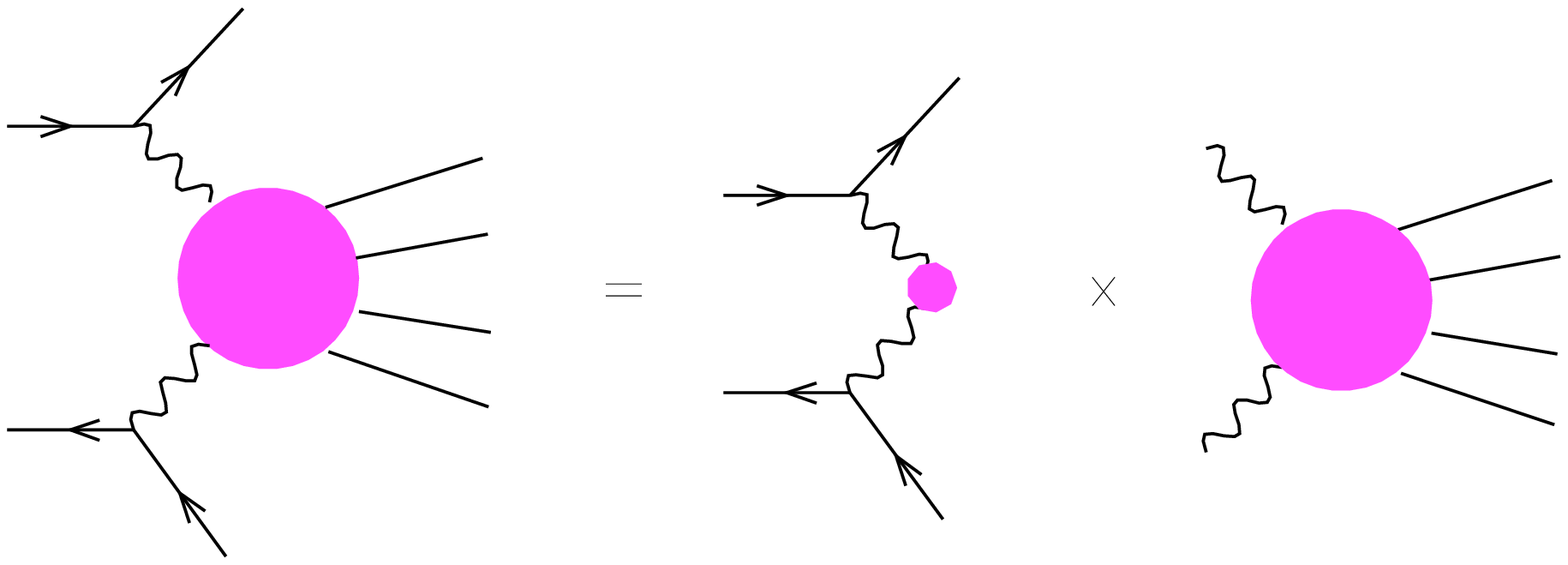,width=0.46\textwidth,clip=}
\caption{Decomposition of the phase space for the process
$\epem\to\epem +hadrons$.}
\label{fig:decomp}
\end{figure}
Having the matrix elements at disposal, one can plug them is his
preferred NLO algorithm, and get physical results. Our case can
however be greatly simplified in a preliminary stage; in fact, the
incoming and outgoing leptons actually do not participate in the hard
scattering, that is initiated by the two virtual photons. Formally,
the simplification goes through a suitable decomposition of the phase
space, which is pictorially represented in fig.~\ref{fig:decomp}. We
write the phase space of two leptons plus $n$ partons (in our case,
$n=2$ or $n=3$ for the one-loop or the tree-level amplitudes
respectively) as follows
\beqn
&&d\Phi_{2+n}(p_1+p_2;p_1^\prime,p_2^\prime,k_1,\ldots,k_n)=
\nonumber \\&&\phantom{aaaa}
d\Gamma(p_1^\prime,p_2^\prime) d\Phi_n(q_1+q_2;k_1,\ldots,k_n),
\label{phspdec}
\eeqn
where $p_i$ ($p_i^\prime$) and $q_i=p_i-p_i^\prime$ are the momenta of 
the incoming (outgoing) leptons and photons respectively, and $k_i$ are 
the momenta of the final state (coloured) partons. The first entry in 
$d\Phi_{2+n}$ and $d\Phi_n$ is the sum of the momenta of the incoming 
particles (the $\epem$ pair and the two-photon pair respectively, as shown
in fig.~\ref{fig:decomp}). Eq.~(\ref{phspdec}) implicitly defines 
$d\Gamma$, since we know {\it a priori} how to write the phase 
space for an arbitrary number of particles, given their momenta. 
We can thus immediately get
\beq
d\Gamma=\frac{d^3 p_1^\prime}{(2\pi)^3 2p_1^{\prime 0}}
\frac{d^3 p_2^\prime}{(2\pi)^3 2p_2^{\prime 0}}.
\label{dGammadef}
\eeq
We have therefore decomposed the original phase space into the product
of two pieces, each of which is Lorentz invariant. We exploit this
fact by re-writing eq.~(\ref{dGammadef}) in the center-of-mass frame
of the incoming $\epem$ pair: defining $S=(p_1+p_2)^2$, we get
\beq
d\Gamma=\frac{1}{4(2\pi)^6 S}dQ_1^2 dQ_2^2 dE_1 dE_2 d\varphi d\bar{\varphi},
\label{dGammaepem}
\eeq
where $\varphi$ and $\bar{\varphi}$ are two generic azimuthal angles, one
of which (let's say $\varphi$ to be definite) can be interpreted as the
angle between the two outgoing leptons; $E_i$ are the energies
of the outgoing leptons in the center-of-mass frame of the incoming $\epem$ 
pair. The strategy of the computation should now be clear. The hard
process that we actually deal with at NLO is that of eq.~(\ref{processgg});
thanks to the decomposition in eq.~(\ref{phspdec}), we have a $2\to n$
phase space which is formally identical to that one gets as a starting
point of any NLO algorithm; thus, we can safely adopt one of the
existing NLO algorithms, and study the process of eq.~(\ref{processgg})
in the $\gamma^*\gamma^*$ center-of-mass frame, without any reference
to the incoming or outgoing leptons: this amounts to a non-trivial
simplification, since the complexity of the numerical computations
at NLO is known to grow fast with the number of particles involved
in the hard scattering. Of course, the information on the lepton
momenta is entering somewhere, in particular in the matrix elements;
to take into account this fact, we proceed in two steps, still using 
fig.~\ref{fig:decomp} as a guide. We start by generating the full kinematical
configuration of the outgoing leptons, using eq.~(\ref{dGammaepem}).
In doing this, we also get the photon momenta, and therefore we know
how to boost from the $\epem$ to the $\gamma^*\gamma^*$ center-of-mass
frame. Then, we boost the lepton momenta to the $\gamma^*\gamma^*$ 
center-of-mass frame, where we generate the remaining (parton) momenta,
according to the phase space $d\Phi_n$; at this point, we can perform
all the manipulations required by the NLO algorithm.

\section{Results}

Following the procedure outlined in the previous section, we constructed
a user-friendly code capable of predicting, to NLO accuracy, any
infrared-safe quantity constructed with up to three partons (plus two
leptons) in the final state.\footnote{The code can be obtained upon 
request by mailing to Stefano.Frixione@cern.ch.} The code is based upon
the NLO algorithm of refs.~\cite{FKS,Jets97}, and it is a suitable
modification of one of the codes presented in ref.~\cite{Jets97}.

In order to produce phenomenological results, we used the following
input parameters, taken from an analysis performed by L3~\cite{Data};
we stress that other physically sensible sets of parameters would lead
to the same qualitative conclusions. The center-of-mass energy is
fixed to $\sqrt{S}=200$~GeV; the scattering angles $\theta_i$ and
energies $E_i$ of the outgoing leptons are required to fulfill 
\mbox{$0.03\le\theta_i\le 0.066$} and \mbox{$E_i>40$~GeV},
respectively (actually, L3 use \mbox{$E_i>30$~GeV}; we prefer a slightly
larger value in order to have larger virtualities. The relative 
difference between the two choices is of the order of
1\%). We use a two-loop expression for $\as$, with
$\as(M_{\sss Z})=0.1181$~\cite{PDG00}. A choice has also to be made
for the scales entering the strong and electromagnetic running
coupling constants.  For the latter, we left the possibility open in
the code to have independent scales for the two photon legs; it seems
therefore natural to choose each scale equal to the virtuality of the
corresponding leg. As far as the scale entering the strong coupling is
concerned, we have chosen
\beq
\mu^2=\frac{Q_1^2+Q_2^2}{2}+
\left(\frac{\sum_{i=1}^n k_{{\sss T}i}}{2}\right)^2,
\eeq
where $k_{{\sss T}i}$ are the transverse momenta (in the $\gamma^*\gamma^*$
center-of-mass frame) of the emitted coloured partons; a discussion
on this choice will be given in ref.~\cite{CDFT}.

\begin{figure}[thb]
\vspace{9pt}
\epsfig{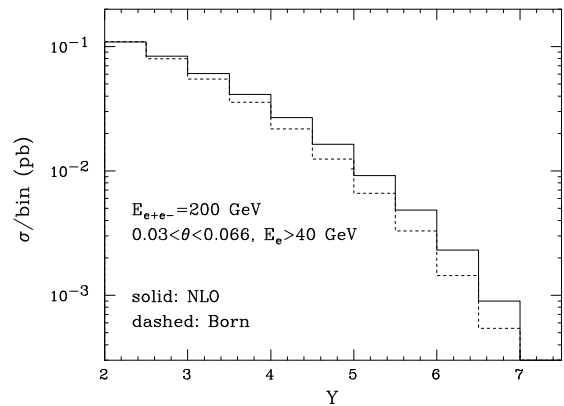}
\caption{Cross section as a function of $Y$. The LO and NLO
results are both shown.}
\label{fig:Yplot}
\end{figure}

In fig.~\ref{fig:Yplot} we present our results for the total
cross section in $\epem$ collisions (that is, relevant to 
the process in eq.~(\ref{processee})), as a function of $Y$. 
This quantity is defined as follows:
\beq
Y=\log(S/S_0),\;\;\;\;S_0=\frac{\sqrt{Q_1^2 Q_2^2}}{y_1 y_2},
\eeq
where 
\beq
y_i=1-\frac{2E_i}{\sqrt{S}}\cos^2(\theta_i/2).
\eeq
It is worth noticing that $Y$ is directly related to the BFKL
logarithm $L$ entering eq.~(\ref{BFKLxsec}). In fact, for large $W$'s,
\mbox{$W\simeq y_1 y_2 S$}; furthermore (see for example ref.~\cite{BHS}),
a sensible choice is \mbox{$\mu_{\sss\rm W}^2=c_{\sss Q}
\sqrt{Q_1^2 Q_2^2}$}, with $c_{\sss Q}$ a suitable constant. It then 
follows that
\beq
L=Y-\log c_{\sss Q}.
\label{LvsY}
\eeq
In fig.~\ref{fig:Yplot}, the dashed histogram represents the LO
result, while the solid histogram is the full NLO result; thus, the
difference between the two histograms is the order-$\as$ correction
computed in this paper. As can be seen from the figure, the NLO
corrections are basically negligible when small $Y$'s are considered.
On the other hand, in the region of intermediate and large $Y$, the
NLO corrections are definitely more important; at $Y=4$ ($Y=6$), the
LO result is increased by a factor of about 1.2 (1.6).  This is a non
trivial piece of news; in fact, when comparing L3 data to LO
predictions at -- say -- $Y=4$, theory underestimates the data by a
factor of about 2.5. This fact is interpreted as the signal of a
dominant contribution coming from diagrams with a $t$-channel gluon
exchange. However, the result given here shows that part of the
discrepancy has to be associated with diagrams that do not have any
$t$-channel gluon exchange. In other words, the BFKL asymptotic region
is probably still far away, and this implies that the contributions of
diagrams with four quarks (that is, of order $\as^2$), but without
$t$-channel gluons, is probably not completely negligible. In any case,
for a precise comparison between theory and data, NLO corrections
cannot be ignored.

\begin{figure}[thb]
\vspace{9pt}
\epsfig{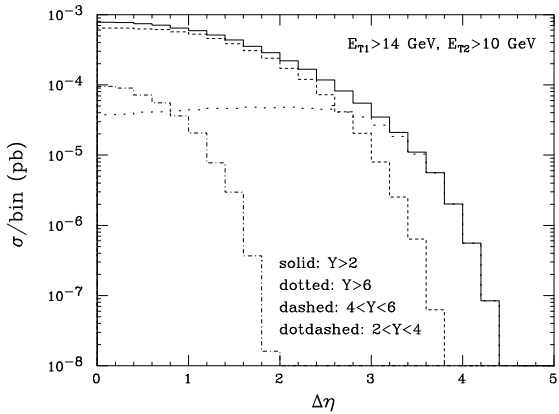}
\caption{Dijet cross section at NLO.}
\label{fig:Deta}
\end{figure}
We now go back to eq.~(\ref{LvsY}), and consider the role of the
constant $c_{\sss Q}$. In ref.~\cite{BHS}, this constant has been
estimated by equating the definition of $L$ with the rapidity difference
between the most forward and backward quarks produced in the hard
scattering. We can repeat the same exercise, using jets instead of
quarks (as one always should, when a NLO computation is available).
We thus computed two-jet cross sections, selecting the most forward/backward
jets and requiring the most (least) energetic of them to have transverse
energy larger than 14 (10) GeV in the center-of-mass frame of the
$\gamma^*\gamma^*$ pair. The difference in their rapidities,
denoted as $\Delta\eta$, is shown in fig.~\ref{fig:Deta}, for various
cuts on $Y$. It is interesting to notice that the contribution due
to the large $Y$ region (dotted histogram, corresponding to $Y>6$) 
basically coincides, in the large $\Delta\eta$ region, with the cross 
section integrated over all $Y$'s. We therefore find what we expect, 
namely that the large $Y$ region is basically populated by events
characterised by two hard jets well separated in rapidity. 
Thus, following ref.~\cite{BHS}, we can invert eq.~(\ref{LvsY}) to
get $c_{\sss Q}$, by substituting $Y=6$ and identifying $L$ with the
average $\Delta\eta$ corresponding to the cut $Y>6$. We have
\beq
c_{\sss Q}=\exp(Y-\langle\Delta\eta\rangle)\simeq 75,
\eeq
which is rather close to the value of 100 estimated in ref.~\cite{BHS}.
\enlargethispage{1000pt}
The author wishes to thank Vittorio Del Duca for reading the manuscript.

\end{document}